%% file: main.tex
\documentclass[12pt,authoryear]{elsarticle}

\usepackage{amssymb}
\usepackage{amsmath}
\usepackage{pgfplots}
\usepackage{multirow}
\usepackage{lscape}
\pgfplotsset{ width =12cm,compat=1.9}
\usepackage{xcolor}
\usepackage{trivfloat}
\trivfloat{quadro}
\definecolor{lightblue}{RGB}{173, 216, 230} 

\usepackage{hyperref}
\hypersetup{breaklinks=true}
\usepackage{tcolorbox}

\journal{Information and Software Technology}

\begin{document}

\begin{frontmatter}


\author{Beatriz Santana\corref{cor1}\fnref{label1}}
 \ead{santanab@ufba.br}
 \cortext[cor1]{Corresponding author}

\title{Psychological Safety in Software Workplaces: A Systematic Literature Review} 

\affiliation[label1]{organization={Federal University of Bahia},
            state={Bahia},
            country={ Brazil}}
 \author[label2]{Lidivânio Monte}
 \ead{lidivanio33@gmail.com}
 \author[label2]{Bianca Santana de Araújo Silva} 
\ead{bianca.silva2344@gmail.com}
 \author[label4]{Glauco Carneiro}
 \ead{glauco.carneiro@dcomp.ufs.br}
 \author[label3]{Sávio Freire}
 \ead{savio.freire@ifce.edu.br}

\author[label2]{José Amancio Macedo Santos} 
\ead{zeamancio@uefs.br}
\author[label1]{Manoel Mendonça} 
\ead{manoel.mendonca@ufba.br}
\affiliation[label2]{organization={State University of Feira de Santana},
            state={Bahia},
            country={Brazil}}
\affiliation[label4]{organization={Federal University of Sergipe},
            state={Sergipe},
            country={Brazil}}
\affiliation[label3]{organization={Federal Institute of Ceará},
             state={Ceará},
             country={Brazil}}



\begin{abstract}
\textbf{Context}: Psychological safety (PS) is an important factor influencing team well-being and performance, particularly in collaborative and dynamic domains such as software development. Despite its acknowledged significance, research on PS within the field of software engineering remains limited. The socio-technical complexities and fast-paced nature of software development present challenges to cultivating PS. To the best of our knowledge, no systematic secondary study has synthesized existing knowledge on PS in the context of software engineering.

\textbf{Objective}: This study aims to systematically review and synthesize the existing body of knowledge on PS in software engineering. Specifically, it seeks to identify the potential antecedents and consequences associated with the presence or absence of PS among individuals involved in the software development process. 

\textbf{Method}: A systematic literature review was conducted, encompassing studies retrieved from four digital libraries. The extracted data were subjected to both quantitative and qualitative analyses.  

\textbf{Results}: The findings indicate a growing academic interest in PS within software engineering, with the majority of studies grounded in Edmondson’s framework. Factors antecedents of PS were identified at the individual, team, and organizational levels, including team autonomy, agile methodologies, and leadership behaviors. The presence of PS was found to positively impact team dynamics, software quality, technical excellence, and job satisfaction. Furthermore, PS was observed to mediate relationships between constructs such as ethical leadership and innovative behavior.  

\textbf{Conclusion}: PS fosters innovation, learning, and team performance within software development. However, significant gaps persist in understanding the contextual factors influencing PS, its underlying mechanisms, and effective strategies for its enhancement. Future research should address these gaps by investigating the practical applications of PS within diverse organizational settings in the software engineering domain.  
\end{abstract}

\begin{highlights}
\item The review revealed an increasing academic interest in psychological safety (PS) within software engineering, with notable publication growth after 2022. Studies are geographically diverse.
\item  Most studies utilized Edmondson’s (1999) concept and questionnaire to investigate PS, highlighting its central role in the field. 
\item Antecedents of PS were identified at individual, team, and organizational levels, such as team autonomy, agile practices, leadership behavior, and cultural differences.
\item PS positively impacted team dynamics, software quality, and individual job satisfaction. However, the consequences of its absence, such as ``brain drain,'' underline the importance of fostering a psychologically safe environment.
\item The mediating role of PS in relationships like ethical leadership \& innovative behavior, and mindfulness \& team performance was also highlighted.
\end{highlights}


\end{frontmatter}




\section{INTRODUCTION}
\label{sec:introduction}

The cultivation of safe and supportive work environments has emerged as a global priority, aligning closely with the objectives outlined in the United Nations Sustainable Development Goals (SDGs) \citep{un_sdg}. Specifically, SDG 8 advocates for inclusive economic growth and the promotion of decent working conditions, while SDG 3 underscores the importance of ensuring health and well-being for all individuals. These objectives highlight the recognition that employee well-being extends beyond physical health to encompass psychological, emotional, and social dimensions.

A critical component of this larger well-being framework is psychological safety (PS), the belief that individuals can express themselves without fear of negative consequences. This concept has been identified as a fundamental factor in enhancing team performance and fostering overall well-being, particularly in high-pressure, innovation-driven environments such as the technology sector \citep{edmondson2018fearless}.

Establishing and maintaining PS in the workplace necessitates an understanding of the environmental, organizational, and individual factors that shape it. Behaviors such as asking questions, admitting mistakes, and offering constructive feedback are essential to fostering PS~\citep{edmondson2018fearless}. Therefore, identifying workplace contexts in which these behaviors are encouraged or suppressed is vital for developing psychologically safe environments.

The significance of PS has been extensively recognized in fields such as healthcare and sports. In hospital environments PS has been linked to patient safety and improved team performance~\citep{o2020systematic}. Similarly, in the domain of sports, research has established a positive association between PS and mental health~\citep{vella2024psychological}. Consequently, substantial research efforts have been dedicated to understanding and fostering PS in these fields. Investigating PS within specific professional contexts contributes to conceptual clarity and facilitates the development of effective strategies for its enhancement~\citep{vella2024psychological}.

Unlike other domains such as healthcare or sports, software engineering (SE) involves highly codified and asynchronous collaborative practices that introduce unique interpersonal dynamics. SE practices, such as code reviews, pull requests, continuous integration, and agile ceremonies, create structured yet evaluative environments where developers must routinely expose their decisions, reasoning, and even mistakes. While these practices are essential for ensuring technical quality and team alignment, they also introduce interpersonal risks that can hinder communication and trust when PS is low~\citep{23}. These risks are further intensified by contextual factors common in SE, including tight deadlines, rapid technological change, and collaboration among professionals with diverse backgrounds. Together, these dynamics can heighten interpersonal tension and complicate efforts to establish and sustain PS. Without inclusive and supportive implementation, standard SE practices may inadvertently undermine it~\citep{23}.

The effects of these practices on PS become particularly evident in daily team routines. For example, code reviews, although essential for maintaining code quality, can become moments of intense scrutiny, where team members fear negative judgment. Similarly, pull requests formalize the process of seeking approval, often making feedback public and persistent in version control histories. Agile ceremonies, such as stand-ups, retrospectives, and sprint planning, require frequent verbal participation and open reflection on team dynamics. When PS is high, these practices can foster learning, collaboration, and innovation. However, in psychologically unsafe environments, they may lead to self-censorship, defensive behaviors, and reduced engagement, ultimately harming both team performance and individual well-being.

Existing research underscores the role of PS in software development. Notably, Project Aristotle, a comprehensive study conducted by Google, identified PS as the most influential determinant of team effectiveness \citep{google_aristotle}. In teams characterized by high PS, members engage more openly in discussions, contribute ideas without fear of ridicule, and embrace a culture that encourages learning from mistakes.

While the benefits of PS on performance and well-being have been extensively documented in domains such as healthcare and sports, its implications in SE remain insufficiently explored. Given the critical role of teamwork, innovation, and collaboration in SE, a deeper understanding of PS within this context is essential. The interdependent nature of SE activities, combined with its technical and interpersonal challenges, renders the field particularly susceptible to adverse effects associated with diminished PS~\citep{21}. Consequently, compromised PS can negatively affect both software quality and team performance.

Although previous studies have examined PS in SE concerning specific outcomes and interactions, no systematic review has been conducted to synthesize existing findings comprehensively. To address this gap, we conducted a systematic literature review to aggregate and analyze the current state of research on PS in the software development context. This study provides an overview of the background, research methodologies, and empirical findings related to PS in SE. Based on the collected evidence, we categorize the findings into two primary domains: antecedents and consequences of PS in SE.

This review identifies key elements that affect (antecedents) and are affected (consequences) by PS, emphasizing the interplay between team practices, organizational contexts, and individual behaviors. Factors such as team performance, autonomy, and social interactions contribute to the establishment of a psychologically safe work environment. Additionally, organizational strategies, including high-commitment work systems and agile methodologies, have been identified as instrumental in maintaining PS within SE teams.

The remainder of this paper is structured as follows: Section~\ref{sec:background} provides a comprehensive overview of PS, drawing upon research from general studies, systematic reviews in other fields, and existing investigations on PS in SE. Section~\ref{sec:researchmethod} outlines the research methodology, detailing the selection and analysis processes. Section~\ref{sec:results} presents the results of the systematic review based on the application of selection criteria to 28 studies. Section~\ref{sec:discussion} discusses the implications of PS within SE. Section~\ref{sec:threatstovalidity} addresses potential threats to the validity of the study. Finally, Section~\ref{sec:conclusion} summarizes the key findings, highlights critical insights, and suggests directions for future research.

\section{BACKGROUND}
\label{sec:background}

This section examines PS, its conceptual evolution, and its relevance across various domains, including organizational settings, healthcare, sports, and SE. The discussion begins with a definition of PS, highlighting its foundational theories and key contributions. Subsequently, it explores studies that identify the antecedents, outcomes, and interventions associated with PS in diverse contexts. Finally, the section narrows its focus to the role of PS in SE, emphasizing its impact on team dynamics and software quality.

\subsection{Definition and Evolution of Psychological Safety}

The concept of PS was first introduced by Schein and Bennis in 1965 as a construct that facilitates adaptation to organizational changes. They argued that individuals in psychologically safe environments feel empowered to modify their behaviors without fear of retaliation or blame, thereby enhancing their ability to address organizational challenges and changes~\citep{schein1965personal}.

Expanding on this perspective, \citet{kahn1990psychological} investigated the relationship between PS and personal engagement at work, analyzing the behaviors of architects and camp counselors in environments perceived as psychologically safe. His findings indicated that PS increases individuals' willingness to express themselves while reducing defensive behaviors. \citet[pp. 9]{edmondson1999psychological} further refined the concept, defining PS within team settings as ``\textit{a shared belief that the team is safe for interpersonal risk-taking}.'' She developed a questionnaire to measure PS, distinguishing it from team cohesion and emphasizing its role in fostering open communication and the freedom to speak without fear of reprisal.

Recognizing the significance of PS in organizational contexts, numerous studies have explored strategies to cultivate PS and examined its associated outcomes. In a meta-analytic review, \citet{frazier2017psychological} analyzed PS across multiple industries, identifying its antecedents and effects. Their findings revealed several factors that contribute to the development of PS, including proactive personality, emotional stability, learning orientation, inclusive leadership, transformational leadership, trust in leadership, work design characteristics, autonomy, interdependence, role clarity, and supportive workplace environments. Of these, interdependence and peer support demonstrated the strongest statistical associations with PS. The study also identified key outcomes positively correlated with PS, such as increased engagement, task performance, information sharing, citizenship behaviors, creativity, learning behaviors, commitment, and job satisfaction, with information sharing and learning behaviors exhibiting the strongest relationships.

Beyond general workplace studies, systematic reviews have been conducted to examine PS in specific domains. For instance, a research \cite{o2020systematic} in the healthcare sector has explored factors that facilitate PS and evaluated interventions designed to improve PS. Similarly, in sports, systematic reviews have sought to define PS, identify its antecedents and attributes, and assess its effects on performance and well-being \citep{vella2024psychological}.

\citet{edmondson2014psychological} highlighted that PS is context-specific, with its formation and maintenance influenced by factors such as team composition, task interdependence, and the nature of the work being performed. Since different domains exhibit unique coordination, communication, and knowledge-sharing patterns, caution is necessary when making empirical generalizations across them.

The SE domain, in particular, has distinct sociotechnical characteristics that shape how interpersonal risk is experienced and mitigated \citep{Mohanani}. Specific practices in software teams, like submitting code for review, suggesting design changes, or refactoring a colleague’s work, influence interpersonal dynamics \citep{23}. These practices demand not only technical proficiency but also social sensitivity, as they expose individual contributions to peer scrutiny and feedback. For example, \citet{Tamburri} introduced the concept of social debt to describe the negative consequences of poor communication, unresolved interpersonal tensions, and misaligned expectations within software teams. The authors argued that many problems perceived as technical are, in fact, rooted in social dynamics, highlighting the importance of understanding and addressing interpersonal relationships within software teams.

The complexity in SE extends beyond technical challenges, such as legacy systems, continuous integration, or the coexistence of multiple programming paradigms, and includes aspects like dependencies among tools, teams, and evolving requirements. Moreover, the constant need for adaptation to new technologies, fluctuating project scopes, and time pressures amplifies the psychological cost of speaking up, admitting mistakes, or challenging design decisions. These behaviors, however, are precisely the ones that underpin a psychologically safe environment.

Therefore, the psychological and social challenges faced by software practitioners, exacerbated by the domain's complexity, rapid pace, and emphasis on collaborative problem-solving, underscore the importance of examining PS as a domain-specific phenomenon.

\subsection{Psychological Safety in Software Engineering}

In the domain of SE, both social and technical factors are critical to project success~\citep{caballero2023community}. PS has been identified as a key determinant of effective teamwork and is influenced by suboptimal socio-technical decisions, commonly referred to as ``community smells''~\citep{caballero2023community}.

Recent studies have increasingly focused on the role of PS in SE, emphasizing its impact on team dynamics and software outcomes. Acknowledging the social context within software development is essential for addressing fundamental human needs, including PS, as these factors directly affect software quality~\citep{13}. Additionally, \citet{10} demonstrated that the level of PS within a team influences members' participation in agile practices, such as daily standup meetings, and fosters behaviors that support continuous software improvement.

Furthermore, \citet{04} highlighted the long-term benefits of PS, emphasizing its role in promoting a culture of learning that enhances a team's capacity for achieving technical excellence. PS enables developers to engage in open discussions regarding process and behavioral failures, propose innovative ideas, and contribute to the continuous improvement of software quality. However, it is imperative to raise awareness about PS within development teams. \citet{07} suggested that intervention activities designed to facilitate open discussions on typically unspoken topics can significantly improve team members' perceptions of PS.

The precursors of PS in agile development have also been examined, with a particular emphasis on fostering psychologically safe environments to maximize the benefits of agile methodologies~\citep{21}. \citet{21} identified leadership behavior as a fundamental determinant of PS, as leaders play a pivotal role in cultivating trust and encouraging team members to express themselves without fear of negative consequences. Moreover, interpersonal trust and a learning-oriented culture were found to be essential for PS, as they promote knowledge-sharing and collaboration within teams.

Despite the recognized benefits of PS in SE, no prior study has systematically synthesized the available scientific evidence on this topic. To address this gap, the present review consolidates existing knowledge on the factors influencing PS, its impact on both social and technical aspects of software development, and the practices that either enhance or inhibit PS within teams.

\section{RESEARCH METHOD}
\label{sec:researchmethod}
To conduct the literature review on PS, we followed \citet{kitchenham2007guidelines}'s guidelines for reviewing evidence in the SE field. Our study aims \textbf{to synthesize the existing evidence on PS in SE, focusing on both its antecedents and consequences}. It also reveals the potential research gaps that require further investigation.

To capture the multifaceted nature of PS, we included qualitative, quantitative, and mixed-methods studies. By incorporating a variety of research designs, we aim to provide a more comprehensive understanding of the factors that influence PS within the context of SE. This approach allows us to consider different perspectives and insights, enriching the findings and offering a well-rounded view of the topic. The following sections outline the research questions that guided this study, along with the data collection methods and analytical approaches employed in our investigation. 

\subsection{Research questions} 
To guide our investigation into PS within the context of software development, we formulate a set of research questions (RQs). These questions aim to explore the key dimensions of PS, the methodologies employed in its study, the geographic and institutional distribution of researchers in the field, as well as the instruments and conceptual frameworks utilized in existing literature. The research questions are detailed below.

\textbf{RQ1: Where and when is the research on psychological safety in software development being conducted?} \textit{Rationale}: Examining the geographic distribution and temporal trends of PS research in software development is crucial for identifying regions leading advancements in the field. This analysis helps uncover potential gaps in coverage, highlights regional research hubs, and provides a historical perspective on the evolution of the topic.

\textbf{RQ2: What concepts and instruments are used in studies on psychological safety?} \textit{Rationale}: Variability in the concepts and instruments used to define and measure PS can impact the comparability of studies. A comprehensive understanding of these definitions and measurement tools is essential for ensuring consistency in future research, enhancing the reliability of findings, and facilitating the practical application of results across diverse contexts.

\textbf{RQ3: What antecedents and consequences of psychological safety have been reported in software development?} \textit{Rationale}: PS manifests in various ways, shaping team interactions, communication, and innovation while being influenced by contextual factors \citep{edmondson2018fearless}. Gaining insight into how PS is perceived in software development is essential for identifying the key factors that foster a safer and more collaborative work environment, thereby providing a foundation for targeted interventions and improvements.

\subsection{Article Search Strategy}
We conducted an automated search across four major online databases: ACM Digital Library, IEEE Xplore, Web of Science, and Scopus. The search query employed the term ``psychological* safe,'' drawing upon search strategies used in prior systematic reviews on PS in other domains~\citep{edmondson2014psychological, frazier2017psychological, newman2017psychological}. The inclusion of the asterisk (*) allowed for variations of the term ``psychological safety,'' such as ``psychologically safe,'' as well as related expressions that may appear in the technical literature.

Given the multidisciplinary scope of Scopus and Web of Science, we restricted the search to article titles, abstracts, and keywords. Additionally, we incorporated terms related to software development (``program* '' OR ``software'') to refine the scope of relevant studies. To further enhance specificity within Scopus, we applied an additional filter to limit results to the field of computer science. The search was conducted in June 2024. Table \ref{table2} presents the detailed search strings used for each digital library.

\begin{table}[t]
\caption{Search string}
\centering
\resizebox{0.95\columnwidth}{!}{%
\begin{tabular}{|l|l|l|ll}
\cline{1-3}
\textbf{Base}                                             & \textbf{Search string}                                                                                                                                                                                                                                    & \textbf{Search based on}                                                                                                                                  &  &  \\ \cline{1-3}
ACM                                                       & ``psychological safe'' OR ``psychologically safe''                                                                                                                                                                                                                                  & Keywords anywhere                                                                                                                                         &  &  \\ \cline{1-3}
IEEE                                                      & (``psychological* safe'')                                                                                                                                                                                                                                  & \begin{tabular}[c]{@{}l@{}}Keywords in full text\\ and metadata\end{tabular}                                                                              &  &  \\ \cline{1-3}
Scopus                                                    & \begin{tabular}[c]{@{}l@{}}1)TITLE-ABS-KEY(``psychological* safe'') \\ AND (LIMIT-TO(SUBJAREA,``COMP''))\\ 2)TITLE-ABS-KEY(``psychological* safe'') \\ AND (TITLE-ABS-KEY(``program*'') OR\\  TITLE-ABS-KEY(``software''))\end{tabular}                           & \begin{tabular}[c]{@{}l@{}}1)Keywords in titles, \\ abstracts, keywords, \\ and research area\\ 2)Keywords in titles, \\ abstracts, keywords\end{tabular} &  &  \\ \cline{1-3}
\begin{tabular}[c]{@{}l@{}}Web of \\ Science\end{tabular} & \begin{tabular}[c]{@{}l@{}}((TI=( (``psychological* safe'') \\ AND (``program*'' OR ``software''))) OR \\ AB=( (``psychological* safe'') AND \\ (``program*'' OR ``software''))) OR\\ KP=( (``psychological* safe'') AND \\ (``program*'' OR ``software''))\end{tabular} & \begin{tabular}[c]{@{}l@{}}Keywords in titles,\\  abstracts, and\\  keywords\end{tabular}                                                                 &  &  \\ \cline{1-3}
\end{tabular}
}
\label{table2}
\end{table}

In addition, we employed a snowballing search, which yielded several new articles for analysis. Snowballing is a method for identifying relevant sources by examining the references of discovered articles~\citep{wohlin2014guidelines}. This technique was applied to expand the set of primary studies for our research. We conducted both forward snowballing (exploring the references cited by the primary studies) and backward snowballing (exploring new studies that cite the primary studies).

\subsubsection{Article selection process}
In our systematic review, two researchers independently conducted the article selection process, adhering to predefined inclusion and exclusion criteria, as shown in Table~\ref{table1}. This dual-review approach aimed to enhance the rigor and reliability of the selection process. Initially, titles and abstracts were screened to eliminate false positives, followed by a thorough full-text review of the remaining articles.

\begin{table}[t]
\centering
\caption{Inclusion and exclusion criteria}
\resizebox{0.99\columnwidth}{!}{%
\begin{tabular}{|l|l|lll}
\cline{1-2}
\textbf{Criterion} & \textbf{\begin{tabular}[c]{@{}l@{}}Description of Inclusion Criterion (IC) / \\ Exclusion Criterion (EC)\end{tabular}}                             &  &  &  \\ \cline{1-2}
IC1                & \begin{tabular}[c]{@{}l@{}}The study presents an investigation of psychological safety\\ within the context of software development.\end{tabular} &  &  &  \\ \cline{1-2}
EC1                & \begin{tabular}[c]{@{}l@{}}The article is complete, thus opinions, keynotes,\\ viewpoints, and posters were excluded from the sample.\end{tabular} &  &  &  \\ \cline{1-2}
EC2                & The full text of the article is available.                                                                                                         &  &  &  \\ \cline{1-2}
EC3                & The article is not written in English.                                                                                                             &  &  &  \\ \cline{1-2}
EC5                & The full text is not accessible.                                                                                                                   &  &  &  \\ \cline{1-2}
EC6                & The article is a short paper with fewer than six pages.                                                                                   &  &  &  \\ \cline{1-2}
EC7                & The article is a secondary study.                                                                                                  
     &  &  &  \\ \cline{1-2}
EC8               & The article is a grey literature (e.g., dissertations, technical reports).                                                                                                                 &  &  &  \\ \cline{1-2}
\end{tabular}
}
\label{table1}
\end{table}

The researchers convened to resolve any discrepancies in the article selection process, ruling out the need for the participation of a third researcher. To assess the original inter-rater reliability of the article selection, we employed Cohen's Kappa coefficient~\citep{cohen1960coefficient}, which yielded a value of 0.86. This indicates near-perfect agreement between the researchers, demonstrating strong alignment in their judgments, consistent with established standards for interpreting Kappa scores~\citep{cohen1960coefficient}. Furthermore, this Kappa value exceeds 0.79, a threshold indicating high agreement in the SE field~\citep{liu2020}.

During the initial screening process, we eliminated several false positives by reviewing the titles and abstracts of the papers. For instance, the article titled ``Gratitude and Adolescents’ Mental Health and Well-Being: Effects and Gender Differences for a Positive Social Media Intervention in High Schools\footnote{https://psycnet.apa.org/record/2023-59338-068}'' was excluded as it focused on a school environment, which falls outside the scope of PS in software development. Additionally, we identified studies that examined PS in contexts such as human-robot interaction and manufacturing. While these articles offered valuable insights, they were deemed irrelevant to the specific focus of our research. In summary, we included 20 articles from a total of 1,024 retrieved from the digital libraries.

\begin{figure}[t]
\centering
\includegraphics[scale=0.7]{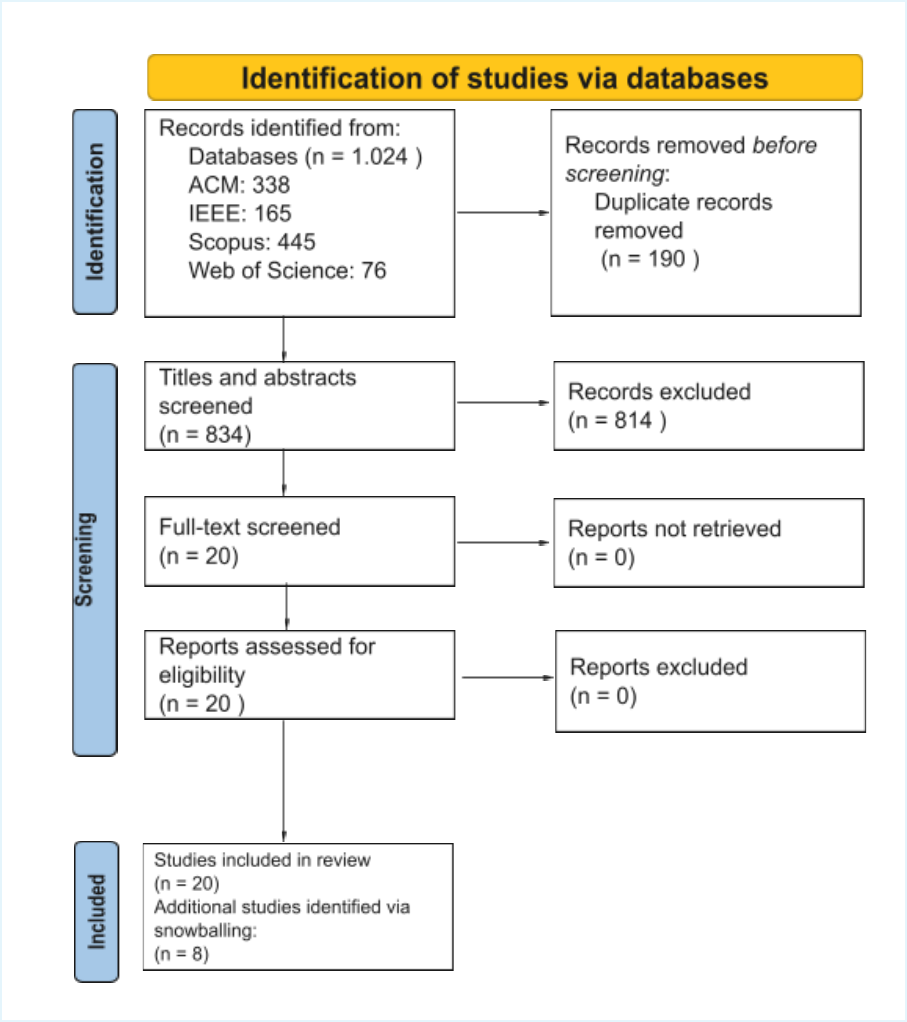}
\caption{Steps to perform the review}\label{fig1}
\end{figure}

The snowballing approach yielded 305 additional articles for analysis. We applied the same selection process to these articles, screening their titles and abstracts, followed by full-text reading of the selected candidates for quality assessment. This process led to the inclusion of 8 additional studies, bringing the total to \textbf{28 articles for analysis}. The selection process is illustrated in Figure~\ref{fig1} based on PRISMA flow diagram\footnote{\url{https://www.prisma-statement.org/prisma-2020-flow-diagram}}.

\subsection{Methodological quality assessment}

We assessed the methodological quality of the included studies using the Mixed Methods Appraisal Tool (MMAT, 2018 version) \citet{hong2018mixed}. The tool includes two initial screening questions and five criteria specific to each methodological category (qualitative, quantitative, or mixed methods).

All studies included in our review successfully met the two screening questions, confirming their eligibility for full appraisal. Regarding the methodological criteria, all studies fulfilled the maximum number of applicable criteria for their category. For studies assessed using five criteria (e.g., qualitative and mixed designs), scores ranged from four to five. For studies assessed using only three criteria—specifically, the 12 non-randomized quantitative studies that did not involve an intervention or exposure—scores were consistently three, which was the highest possible for that category. This adaptation follows MMAT’s guidance for applying criteria flexibly based on study design.

These results indicate a generally high level of methodological rigor across the included studies. A detailed breakdown of individual ratings is available in the supplementary material.

\subsubsection{Data collection}

Considering the collection of 28 articles\footnote{The complete list of the articles is available in~\ref{appendix:list}.}, two researchers independently examined these articles to identify and extract pertinent data, adhering to the attributes specified in Table~\ref{table:attributes}. A description of each of these attributes is also provided in the table.

\begin{table}[]
\centering
\caption{Extracted attributes}
\resizebox{0.9\columnwidth}{!}{%
\begin{tabular}{lllll}
\cline{1-3}
\multicolumn{1}{|l|}{\textbf{Attribute}} & \multicolumn{1}{l|}{\textbf{Description}} & \multicolumn{1}{l|}{\textbf{RQ}} &  &  \\ \cline{1-3}
\multicolumn{1}{|l|}{Title} & \multicolumn{1}{l|}{Title of the selected article} & \multicolumn{1}{l|}{-} &  &  \\ \cline{1-3}
\multicolumn{1}{|l|}{Year} & \multicolumn{1}{l|}{Year of publication of the selected article} & \multicolumn{1}{l|}{RQ1} &  &  \\ \cline{1-3}
\multicolumn{1}{|l|}{Country} & \multicolumn{1}{l|}{Country of publication of the article} & \multicolumn{1}{l|}{RQ1} &  &  \\ \cline{1-3}
\multicolumn{1}{|l|}{PS measures} & \multicolumn{1}{l|}{Which instruments were used to analyze PS} & \multicolumn{1}{l|}{RQ2} &  &  \\ \cline{1-3}
\multicolumn{1}{|l|}{Identification of PS} & \multicolumn{1}{l|}{What concepts were used to characterize PS} & \multicolumn{1}{l|}{RQ2} &  &  \\ \cline{1-3}
\multicolumn{1}{|l|}{Results related to PS} & \multicolumn{1}{l|}{What results were reported in terms of PS} & \multicolumn{1}{l|}{RQ3} &  &  \\ \cline{1-3}
 &  &  &  &  \\
 &  &  &  &  \\
 &  &  &  &  \\
 &  &  &  &  \\
 &  &  &  & 
\end{tabular}%
}
\label{table:attributes}
\end{table}

\subsection{Data Analysis}

This review employed descriptive statistics to analyze the year and country of publication (RQ1). For the publication year, we recorded the year of each study's publication to observe trends in research over time. For the country of publication, we classified the studies by their respective countries to understand geographical trends in research on PS in software development.

For the open data analysis (RQ2 and RQ3), we applied a specific coding process. When categorizing PS measures (RQ2), we identified the data collection instruments used and the authors who defined them. For example, the following passages—``\textit{Employees rated their psychological safety perceptions using a five-item measure from Liang et al. (2012)\footnote{{\citep{12}}}}'' and ``\textit{Team psychological safety (PS) was adapted from Edmondson [20]; a questionnaire-based survey was conducted to test the proposed research model}\footnote{{\citep{18}}}''—indicate that Liang et al. and Edmondson were the authors of the respective instruments. These instruments were then coded accordingly.

Similarly, to identify the PS concept (RQ2), we examined the authors' definitions and statements about PS in each study. For instance, in the following passage: \textit{``In 1999, Edmondson published her seminal work on psychological safety [6], defining the term as a shared belief held by members of a team that the team is safe for interpersonal risk-taking\footnote{{\citep{07}}}.''} we collected this concept by contrasting it with definitions in other studies considered in this review.

Regarding how PS is perceived in software development (RQ3), we analyzed the variables described in the selected studies as an antecedent or consequence of PS. We then employed the following categorization process:

\begin{itemize}
\item \textbf{Antecedents of PS:} This category highlights how factors such as spontaneous interaction can influence PS. For instance, one study reported that ``\textit{Psychological safety is enabled by spontaneous interaction}\footnote{{\citep{02}}}.'' 

\item \textbf{Consequences of PS:} This category describes how PS can influence factors such as team performance. For example, some studies indicated that ``\textit{Team performance and individual job satisfaction are predicted by psychological safety}\footnote{{\citep{01}}}.'' 
\end{itemize}

This categorization facilitated a detailed analysis of how PS interacts with different factors.

\subsection{Data Synthesis}

We employed a narrative synthesis approach \citep{popay2006guidance} to organize and interpret the findings of the included studies, due to the heterogeneity in research aims, contexts, and methodological choices. The synthesis process integrated the analytical strategies used to address the three research questions of this review.

For RQ1, which examined when and where the selected studies were published, we extracted and described the year and country of publication of each study. This information was used to identify temporal and geographical trends in PS research in SE.

For RQ2, which focused on how PS is measured and conceptualized, we began by identifying the instruments used across studies and categorizing them according to the author(s) who developed or adapted the measurement. Similarly, we examined the conceptual definitions of PS provided in each study. This allowed us to group studies based on conceptual alignment and instrument adoption.

To address RQ3, which investigates antecedents and blue{consequences} of PS in software development, we synthesized the variables reported across studies as being related to PS. These variables were initially classified according to their relationship with PS — as either antecedents (factors influencing PS) or consequences (factors influenced by PS).

Given the conceptual convergence among several variables across studies, we organized these antecedents and consequences into three broad contextual categories: individual, team, and organizational. This categorization was chosen to accommodate similarities in variable types and to provide a coherent analytical structure. For instance, variables referring to personal attributes were grouped at the individual level, while team dynamics and organizational practices were grouped at their respective levels. This structure allowed us to synthesize patterns across studies despite variations in study design or terminology.

The identification and initial classification of data across all research questions were performed independently by two authors. The final grouping of variables and the narrative synthesis were conducted by the first author and validated by two co-authors to ensure consistency and analytical rigor.

\section{Results}
\label{sec:results}
This section presents the results related to the research questions. The raw data is available in the supplementary material.

\subsection{RQ1: Where and when is the research on psychological safety in software development being conducted?}

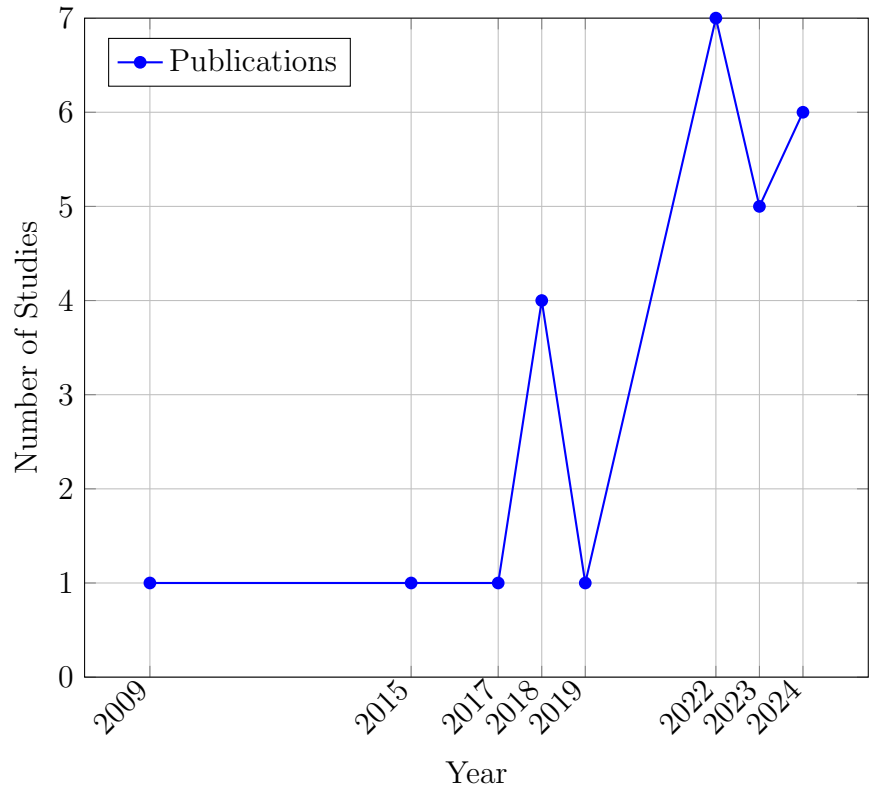
\begin{figure}[!ht] 
\centering
\begin{tikzpicture}
    \begin{axis}[
        xlabel={Year},
        ylabel={Number of Studies},
        xtick=data,
        ymin=0, ymax=7,
        xticklabel style={/pgf/number format/1000 sep=, rotate=45, anchor=east},
        legend pos=north west,
        grid=both
    ]
    \addplot[
        color=blue,
        mark=*,
        thick
    ] coordinates {
        (2009,1) (2015,1) (2017,1) (2018,4) (2019,1) (2022,7) (2023,5) (2024,6)
    };
    \legend{Publications}
    \end{axis}
\end{tikzpicture}
\caption{Trends in number of studies over time}
\label{fig:year}
\end{figure}

Figure~\ref{fig:year} shows the number of studies published on PS in SE over the years. It reveals significant growth in research in this area. Initially, only one article was published in 2009, 2015, 2017, and 2019, but we saw a spike in 2018. Interest in the topic grew from 2022 onwards, the same year we had the most significant increase in publications.

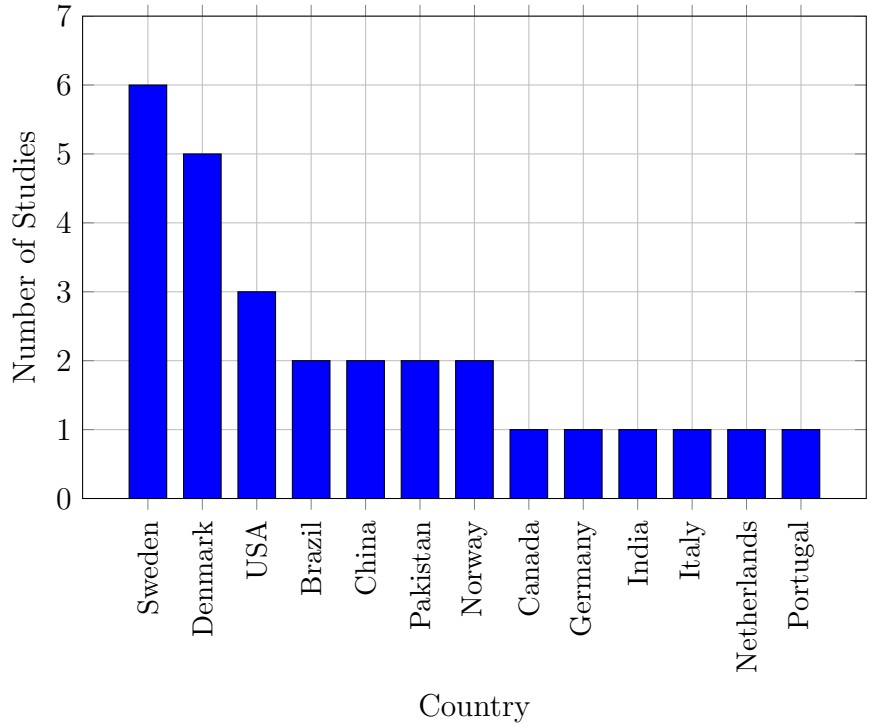
\begin{figure}[!ht] 
\centering
\begin{tikzpicture}
    \begin{axis}[
        ybar,
        ylabel={Number of Studies},
        symbolic x coords={Sweden,Denmark,USA,Brazil,China,Pakistan, Norway, Canada,Germany,India,Italy,Netherlands,Portugal},
        xtick= data,        
        ymin=0, ymax=7,
        width=12cm,
        height=8cm,
        bar width=0.5cm,
        grid=both,
        xlabel={Country},
        ymajorgrids=true,
        xticklabel style={font=\small, rotate=90, anchor=east}
    ]
    \addplot[fill=blue] coordinates {
        (Sweden,6) (Denmark,5) (USA,3) (Brazil,2)
        (China,2) (Pakistan,2) (Norway,2) (Canada,1) (Germany,1)
        (India,1) (Italy,1) (Netherlands,1) (Portugal,1)
    };
    \end{axis}
\end{tikzpicture}
\caption{Geographic Distribution of Studies}
\label{fig:country}
\end{figure}

Figure \ref{fig:country} shows the geographic distribution of studies published on PS in SE. It highlights the diversity of countries involved in researching this topic. Among the listed countries, Sweden stands out with six publications, closely followed by Denmark with five studies. The United States has three publications, while Brazil, China, and Pakistan each have two publications at the time of this review. Additionally, several countries, including Canada, Germany, India, Italy, the Netherlands, and Portugal, have contributed with one publication each.

\begin{tcolorbox}[colback=gray!2, colframe=black, width=\linewidth]
\textbf{Finding \#1:} There has been a notable increase in research on PS within the field of SE over time, with a marked surge in interest in 2022, which saw the highest number of publications to date. Furthermore, there is a broad geographical distribution of countries contributing to PS research. Among the nations identified, Sweden stands out with six publications, closely followed by Denmark with five studies.
\end{tcolorbox}

\subsection{RQ2: What concepts and instruments are used in studies on psychological safety?}
We found that 24 studies adopted the concept of PS as defined by \citet{edmondson1999psychological}. This widespread adoption highlights the significant influence of Edmondson's work in the field, establishing a foundational theoretical framework for understanding the phenomenon. Among these, two studies also incorporated the definition of PS proposed by \citet{kahn1990psychological}. One study relied exclusively on \citet{kahn1990psychological}'s definition, while the remaining studies did not explicitly use a formal concept of PS.

According to \citet[pp. 9]{edmondson1999psychological}, PS is defined as \textit{``a shared belief held by team members that the team is safe for interpersonal risk-taking''}, while \citet[pp. 18]{kahn1990psychological} described it as \textit{``Psychological safety was experienced as feeling able to show and employ one's self without fear of negative consequences to self-image, status, or career. People felt safe when they trusted they would not suffer for their engagement''}. Based on these definitions, we propose the following definition of PS: ``the ability to take interpersonal risks without fear of negative repercussions for one's image, status, or career.'' We believe that this definition encompasses the key concepts proposed by both \citet{edmondson1999psychological} and \citet{kahn1990psychological} in a concise and straightforward manner.

\begin{tcolorbox}[colback=gray!2, colframe=black, width=\linewidth]
\textbf{Finding \#2:} The authors have utilized the PS concepts defined by~\citet{edmondson1999psychological} and~\citet{kahn1990psychological} when examining this construct within the context of SE. We have consolidated these definitions into the following concept:
\\
\\
\textbf{``Psychological safety is the ability to take interpersonal risks without fear of negative repercussions for one's image, status, or career.'' }
\end{tcolorbox}

We examined the instruments employed to investigate PS in software development contexts and found a strong reliance on Edmondson’s team-level scale~\citep{edmondson1999psychological}. Among the 28 studies included in this review, 15 employed this instrument, with nine of them specifically using the original 7-item version measured on Likert scales ranging from 1 to 5 or 1 to 7. Three studies adapted the items to better suit the SE context—for instance, by rephrasing statements to emphasize openness to technical suggestions rather than interpersonal vulnerability~\citep{13,14,18}. Despite these adaptations, internal consistency remained high, with reported Cronbach’s alpha values ranging from 0.82 to 0.90.

In addition to Edmondson’s scale, three studies employed alternative instruments: one used a validated questionnaire based on the dimensions of organizational learning by~\citet{marsick2003demonstrating}, another used items from~\citet{liang2012psychological}, and one adopted the approach proposed by~\citet{o2020measuring}. A total of 17 studies assessed PS through surveys or questionnaires—most of which were derived from or inspired by Edmondson’s work—while nine studies utilized interviews with customized scripts as their primary data collection method. Two studies included discussion analyses, which were also grounded in Edmondson’s and O’Brien et al.’s frameworks~\citep{o2020measuring}.

Building on the observation that a substantial proportion of studies employed surveys or questionnaires, an analysis of the underlying research designs reveals that the majority of empirical investigations into the relationships between psychological safety and other variables adopt a quantitative approach, relying predominantly on single-source data.

This heavy reliance on single-source data raises significant concerns regarding common method bias (CMB) \citep{podsakoff2024common}. The relationships identified in such studies may be artificially inflated due to shared method variance, thereby weakening the strength of causal inferences, even when statistical significance is achieved. Notable examples include studies by \citet{01}, \citet{14}, and \citet{03}, all of which relied exclusively on questionnaires completed by the same respondents. As \citet{01} acknowledge, this methodological choice may inflate correlation coefficients and undermine causal validity. Similarly, \citet{14} recognize that such data are vulnerable to biases and that ``common method bias may inflate the observed relationships," even when statistical techniques such as PLS-SEM are employed.

In contrast, some studies have adopted alternative methodological approaches. For example, \citet{17} employed a mixed-methods design, combining an exploratory qualitative phase involving interviews with a subsequent quantitative survey phase. Although the quantitative phase still relied on single-source data, the authors acknowledged the potential for common method bias and conducted statistical tests to assess its presence. The integration of a qualitative component provided deeper contextual understanding and introduced a distinct data source, thereby enhancing the overall validity of the study.

A further example of a qualitative approach is found in \citet{21} study, which utilized semi-structured interviews with software professionals to investigate the antecedents and consequences of psychological safety in large-scale agile teams. Qualitative research designs are not susceptible to common method bias in the same manner as quantitative single-source approaches, and they offer rich, contextualized insights that contribute to a more comprehensive understanding of the studied relationships.

This systematic review did not identify any studies that utilized multi-source designs (e.g., data collected from different respondents) to investigate psychological safety in software engineering teams. This represents a notable gap in the existing literature, implying that the robustness of evidence regarding causal relationships and direct impacts on performance and outcome variables still lacks validation through methodologies that more effectively mitigate common method bias.

Furthermore, despite the widespread adoption of Edmondson’s scale, the studies varied substantially in their research aims, analytical strategies, and model structures. Several studies employed complex statistical techniques such as structural equation modeling (including reflective, formative, and multilevel models)\citep{03,20,27}, while others utilized hierarchical regression\citep{01} or multilevel path models~\citep{19}. These methodological differences reflect a diversity in how PS is operationalized and explored across SE contexts.

Overall, the extensive use and reliability of Edmondson’s instrument in these studies support its robustness for capturing PS in software teams. Nonetheless, some findings raise important considerations for future research, particularly regarding potential issues with scale dimensionality~\citep{14}, cultural interpretations, and the influence of specific research contexts \citep{13} on how PS is perceived and measured.

\begin{tcolorbox}[colback=gray!2, colframe=black, width=\linewidth]
\textbf{Finding \#3:} The majority of the studies have utilized the questionnaire proposed by~\citet{edmondson1999psychological} to investigate PS in software development workspaces. 
\end{tcolorbox}

\subsection{RQ3: What antecedents and consequences of psychological safety have been reported in software development?}

\begin{figure}[t]
\begin{center}
\hspace{-13mm} \includegraphics[scale=0.28]{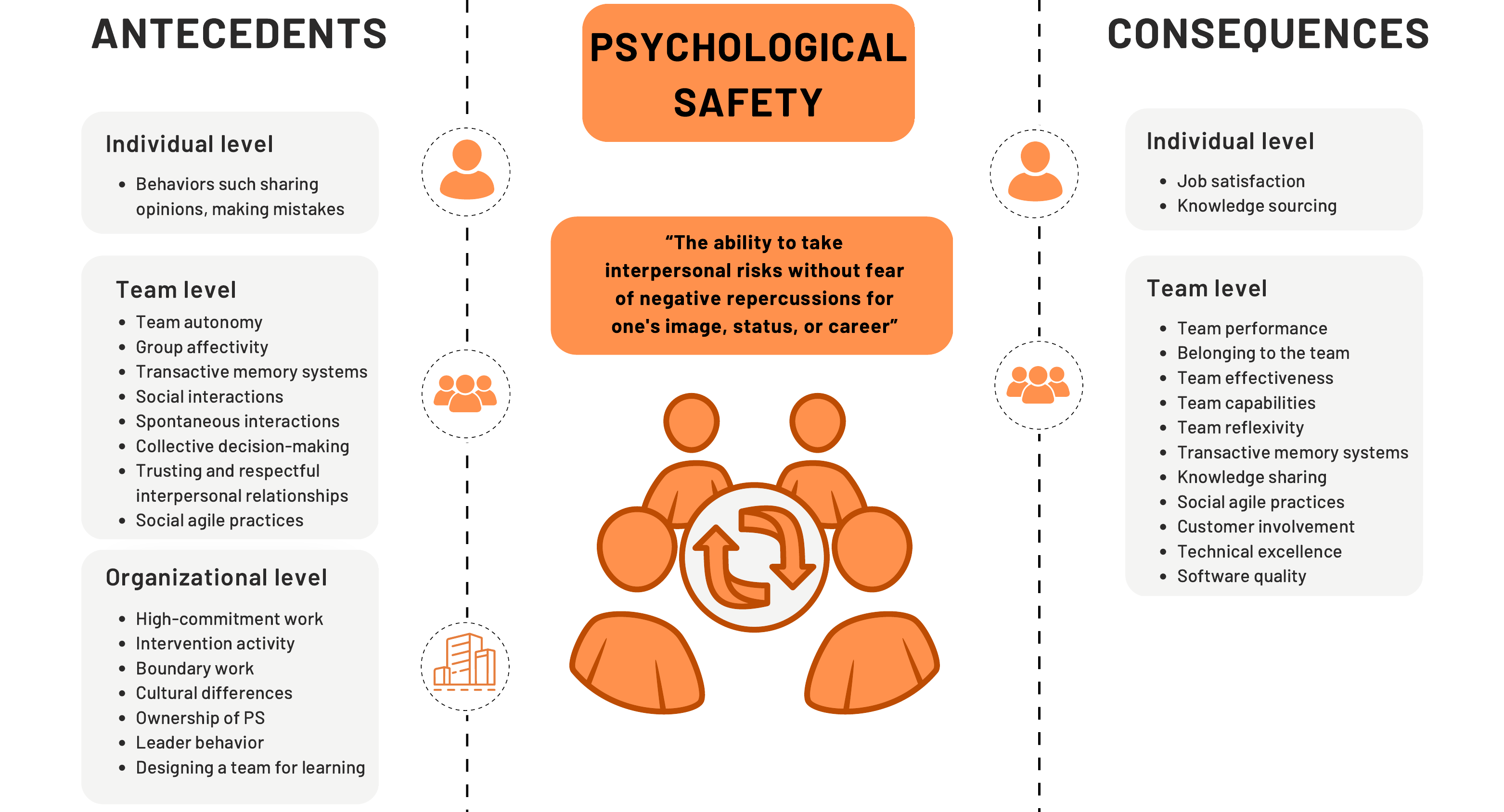}
\end{center}
\caption{Antecedents and consequences of psychological safety have been reported in software development}\label{fig2}
\end{figure}

We identified the interaction between PS and various factors within software development environments. Based on this interaction, we categorized the findings of the studies into two primary categories: antecedents and consequences of PS. The first category includes a range of elements that contribute to the establishment and maintenance of PS within teams, while the second category focuses on the outcomes that PS affects. Figure~\ref{fig2} provides a summary of these categories\footnote{Table~\ref{tab:appendix_Factors_Study} offers a detailed summary of the factors identified in each study.}. The subsequent subsections provide a more in-depth explanation of these categories. 

\subsubsection{Antecedents of psychological safety have been reported in software development}

In relation to the antecedents of PS, we identified both elements that promote PS, such as spontaneous interactions, the use of tools, and agile practices, as well as factors that may undermine it, including high negative affectivity within the group. These factors are organized into three distinct levels: organizational, team, and individual levels, as illustrated in Figure~\ref{fig2}. Below, we provide a detailed explanation of each level.

\begin{itemize} 

\item \textbf{Team level}: We identified seven studies exploring team dynamics concerning PS. \citet{27} noted that team autonomy enhances PS, while~\citet{09} provided evidence supporting the promotion of PS environments based on group effectivity, with negative affectivity contributing to psychologically unsafe environments. Furthermore,~\citet{09} found that transactive memory systems — storing and sharing knowledge within a team — positively impact PS. \citet{02} and \citet{05} examined the effects of remote work on PS. Both studies highlighted that social and spontaneous interactions were critical in influencing PS. The lack of interaction has negative consequences, while the presence of such interactions fosters PS. \citet{17} identified collective decision-making as a behavior that precedes PS at the group level. Furthermore, ~\citet{21} highlighted the fostering of interpersonal relationships of trust and respect as a relevant antecedent for the team. \citet{10} highlighted that social agile practices, including Daily Stand-ups, Reviews and Retrospectives, Pair Programming, Sprint Planning and Prioritization, Collective Code Ownership, and Cross-Functional Teams, positively impact PS, enhancing and strengthening this aspect within the team.

    \item \textbf{Organizational level}: Regarding practices that promote PS at the organizational level,~\citet{12} indicated that high-commitment work systems\footnote{High-commitment work systems encompass a set of systematic human resources practices~\citep{12}.} facilitate PS.  Additionally,~\citet{07} found that using tools as an intervention activity facilitated the enhancement of PS within teams. In~\citet{24}, boundary work is a factor affecting PS, with three specific boundary activities—boundary spanning, buffering, and boundary reinforcement — playing a role. This factor can be moderated by environmental factors such as team task uncertainty and resource scarcity. ~\citet{26} identified cultural differences as challenges to maintaining PS. \citet{17} identified ownership of PS as an antecedent, highlighting the responsibility of organizational leaders in fostering a culture that prioritizes PS. Additionally, ~\citet{21} highlighted leader behavior and designing a team for learning as antecedents that require organizational efforts to establish and sustain practices that enable PS.

    \item \textbf{Individual level}: We identified four studies that examined behaviors that impact PS at the individual level. \citet{17} investigated the antecedents of PS, focusing on strategies to promote this aspect within agile development teams. The other studies highlighted behaviors that indicate a lack of PS~\citep{22}, presented indicators of PS~\citep{06}, and outlined interpersonal challenges and related practices that affect PS~\citep{23}.

    Specifically, \citet{17} identified two antecedents: openness and no blame, which are foundational and at the individual level. \citet{22} observed situations where a lack of PS included divergent ideas, communication issues, and the absence of good programming practices. Moreover, they highlighted that these challenges occurred across various types of relationships, including those with customers, different departments, developers, leaders, and team members.~\cite {23} revealed that interpersonal challenges arising from certain SE practices could undermine PS. Seven specific interpersonal challenges were identified, including disagreeing with suggestions or ideas and calling attention to errors. These challenges can occur in a variety of practices, such as requirements definition, code review, and development process decisions. ~\citet{06}, in their turn, found the presence of specific behaviors — sharing opinions, making mistakes, raising problems, asking questions, and showing consent using online tools — to indicate PS in online retrospectives. 
\end{itemize}

\subsubsection{Consequences of psychological safety have been reported in software development}

A review of the consequences of PS revealed its impact at the individual, and team levels (see Figure~\ref{fig2}). Additionally, we identified aspects directly related to software development, such as software quality, which are affected by PS as well. These factors are discussed below:

\begin{itemize}
    \item \textbf{Team level}: Four studies indicated that PS positively impacts team performance~\citep{01,08,11,27}. Beyond performance, PS is positively associated with other group attributes, such as belonging to the team~\citep{14}, team effectiveness~\citep{15}, team capabilities~\citep{18}, team reflexivity~\citep{27}, transactive memory systems~\citep{09}, and knowledge sharing~\citep{25}.

   Regarding aspects directly related to software development, we found that social agile practices depend on and are influenced by the level of PS~\cite {10}. PS indirectly influences the success of agile projects through its positive impact on customer involvement~\citep{18}. Technical excellence is facilitated by PS~\citep{04}, as it promotes behaviors such as proactive initiative and ongoing learning without fear of consequences.~\citet{13} and~\citet{16} suggested that PS has contributed to software quality by facilitating open discussion of issues and admission of mistakes~\citep{13}, as well as fostering efforts and transparency regarding quality expectations~\citep{16}.
    
    \item \textbf{Individual level}:~\citet{01} suggested that PS positively influences individual job satisfaction. At the same time, the choice of knowledge sources (``knowledge sourcing'') is affected by an individual’s level of PS~\citep{03}. Those with high levels of PS tend to consult team members, whereas those with low levels of PS prefer external sources. 
\end{itemize}

\begin{tcolorbox}[colback=gray!2, colframe=black, width=\linewidth]
\textbf{Finding \#4:} PS is shaped by a variety of antecedents and, in turn, leads to several important consequences in software development contexts. Key antecedents include individual, team, and organizational-level factors such as autonomy, social agile practices, and interpersonal behaviors (e.g., expressing opinions). The consequences of PS involve positive impacts on team performance and software quality, including greater customer engagement and increased transparency in technical discussions.
\end{tcolorbox}

\subsection{Other findings}
In addition to the antecedents and consequences of PS, we identified the results that the PS did not moderate nor was moderated by other variables. For instance, \citet{20} indicated that PS does not moderate the relationship between social anxiety disorder (SAD) and psychological distance. Similarly, \citet{01} found that team norms do not moderate the relationship between team performance and PS.

Regarding indirect effects, \citet{28} found that PS mediates the relationship between ethical leadership and innovative work behavior and the interactive effect between ethical leadership and proactive personality. Similarly, \citet{19} reported significant findings on the mediating role of PS in the relationship between software developers' mindfulness and team performance. Lastly, we identified a study presenting ``brain drain'' as an effect of a lack of PS \citep{21}.

\begin{tcolorbox}[colback=gray!2, colframe=black, width=\linewidth]
\textbf{Finding \#5:} 
   PS may interact with various factors, acting as a mediator in these relationships. Its absence can lead to detrimental effects for organizations. For example, PS serves as a critical link between ethical leadership and innovative work behavior, while also fostering interactive effects between leadership and proactive personality. Additionally, PS mediates the relationship between developers' mindfulness and team performance. The lack of PS can also result in negative outcomes, such as "brain drain," underscoring its significance in organizational settings.
\end{tcolorbox}


\section{Discussion}
\label{sec:discussion}
This section discusses our findings, contrasting them with related work. It also offers implications for researchers and practitioners. 

\subsection{Contrasting with related work}

As stated by~\citet{edmondson2018fearless}, psychologically safe workplaces provide significant advantages in competitive industries. Therefore, understanding how PS is influenced by various factors and its impact on the organizational context can be essential for fostering healthier and more productive work environments. According to the data, we identified that PS is a globally relevant topic. However, with varying levels of attention, the topic is growing in importance within the academic community, as shown in RQ1. 

The limited number of studies on PS in software development highlights the need to investigate the topic. On the other hand, the diverse focus of these studies prevents the generalization of results. For instance, well-established findings on PS's influence on performance and learning in fields like social psychology and healthcare management~\citep{edmondson2014psychological, frazier2017psychological} have not been fully explored in SE contexts. Despite this, some studies have shown that PS positively impacts both team performance and individual learning, suggesting potential benefits for software development environments.

Most of the studies used the concept of PS defined by~\citet{edmondson1999psychological} and~\citet{kahn1990psychological} (RQ2), requiring more investigation on how this concept can be adapted for the context of software development. The primary research method used in the studies was a survey, revealing that longitudinal and interview studies could be applied to investigate the benefits of a psychology safety software workspace. 

The relationship between performance and PS is even more evident when it involves tasks that involve creativity and collaboration~\citep{edmondson2014psychological}, such as software development. In this sense, we identified studies that show the positive effect of performance on PS in software development (RQ3).

Our review reveals that PS is linked to improved team attributes such as belonging, effectiveness, capabilities, and reflexivity. However, the studies did not address factors like geographic dispersion and team diversity~\citep{edmondson2014psychological}, which have been researched in other settings. This omission points to a need for further exploration of how these variables affect PS in software teams.

In terms of learning, previous research has predominantly focused on how PS fosters characteristics such as creativity, error reporting, and knowledge sharing~\citep{edmondson2018fearless}. However, our review found limited exploration of its impact on knowledge sharing, knowledge sources, and transactive memory systems within software development. This suggests a gap in understanding how PS influences learning processes in this field.

At the individual level, only two studies explored the outcomes of PS, focusing on job satisfaction and knowledge sources. Other aspects, such as job engagement, quality internal auditing, learning from failure, and creative work involvement~\citep{edmondson2014psychological}, remain underexplored and warrant further investigation. Besides, while behaviors and cultural differences have been identified as individual-level factors affecting PS, aspects like professional responsibility and individual differences~\citep{o12020systematic} have yet to be examined in this context.

The consequences of lacking PS are severe, ranging from business failures and reduced return on investment to physical harm~\citep{edmondson2018fearless}. For instance, in hospitals, high employee turnover is seen as a significant financial burden due to the costs of training and recruitment, which has driven a focus on PS in that field. However, in software development, only one study specifically examined the effects of a lack of PS \citep{21}.

Behaviors such as expressing opinions, discussing mistakes, and suggesting changes are hallmarks of a psychologically safe environment~\citep{edmondson2018fearless}. Yet, these were identified as challenges in the software development context studied~\citep{23}. This suggests that while PS is essential for practices like agile retrospectives, its absence or difficulty in enacting these behaviors requires deeper investigation. Only one study addressed the nuances of software development processes related to PS, highlighting a significant gap in understanding how specific development environments influence this dynamic \citep{23}.

Our findings also emphasize the critical role of leadership in fostering PS (RQ3). Aspects such as ``Leader behavior'' and ``Ownership of PS'' underscore the importance of leaders in creating supportive environments. Meanwhile, team-level factors like ``Collective decision-making'' and ``Trusting and respectful interpersonal relationships'' suggest that fostering collaboration and mutual respect is vital for building PS.

Regarding organizational antecedents, \citet{newman2017psychological} and \citet{vella2024psychological} identified that factors like supportive practices and organizational support significantly influence PS. While the studies we reviewed did not focus on these organizational antecedents, and practices like high-commitment work systems have been shown to have an impact on PS.

Interventions to enhance PS have been well-studied in healthcare, where they aim to provide practical strategies for establishing and maintaining PS~\citep{o2020systematic}. However, only one study \citep{07} in our review reported using intervention practices, highlighting a need for more research on effective interventions in software development contexts. 

Additionally, aspects such as job commitment, mental health, turnover intentions, and innovation~\citep{edmondson2018fearless} remain underexplored in software development. Understanding these dimensions is important for developing comprehensive strategies to enhance PS and improve team dynamics and software quality.

The results indicate that PS remains an area requiring further understanding within the development context. Questions regarding the extent to which factors influencing PS impact its development and sustainability remain open. Furthermore, understanding how PS affects specific aspects of the development process has yet to be fully addressed. The stages of development where PS can be influenced or cultivated also present areas that warrant further exploration.

\subsection{Distinctive features of psychological safety in software engineering}
While PS has been studied in various domains such as healthcare, education, and manufacturing, our review highlights aspects that make its antecedents and consequences distinctive within the context of SE, as described below:

\begin{itemize}
    \item Software development is characterized by high levels of complexity, uncertainty, and frequent changes in requirements and technologies. These factors amplify the need for teams to communicate openly about errors, raise concerns, and challenge decisions, behaviors strongly linked to PS. However, these same dynamics also increase interpersonal risks, making PS more fragile and difficult to sustain.
    \item Some antecedents of PS in SE are closely tied to technical practices and roles unique to this field. For example, our review identified that practices such as code review, architectural decisions, and task estimation frequently surface as moments of interpersonal tension \citep{23}. These practices often require critical feedback, negotiation, and error disclosure in front of peers, and may involve power asymmetries between developers, team leads, and managers \citep{edmondson2018fearless, 23}.
    \item The consequences of PS in SE extend beyond team satisfaction and learning behaviors. Our findings suggest that PS is linked to software quality \citep{13,16}, technical excellence \citep{04}, customer involvement \citep{18}, and  success of agile project \citep{18}. These outcomes are particularly salient in SE, where both the process and the product are knowledge-intensive and heavily reliant on collaboration.
    \item The distributed and asynchronous nature of many software teams, often using digital tools for collaboration, introduce additional nuances to how PS is built and maintained. Unlike co-located teams in other industries, software teams may have fewer informal interactions to repair ruptures in trust or clarify misunderstandings, increasing the relevance of intentional practices to foster PS \citep{02,05}.
\end{itemize}

In summary, these distinctions suggest that, while foundational theories of PS are applicable, research and interventions in SE must account for the unique sociotechnical dynamics of the field.

\subsection{Implications for researchers and practitioners}
This work highlights several gaps that deserve attention in future research. One key area is the need to explore further the effects of PS in specific software development contexts, such as methodologies and work practices, and in aspects such as technical debt and code smells. Another direction is the use of longitudinal approaches, which can provide a broader view of the evolution of PS and its relationship with team performance and software quality.

In addition, it is essential to broaden the scope of studies to include variables such as cultural differences, the impact of different leadership styles, and the relationship between PS and developers’ mental health. Finally, research on interventions that promote PS, especially in technical teams, can yield important practical insights.

For practitioners, the findings reinforce the importance of creating and maintaining psychologically safe environments in software development teams. Leaders and managers should adopt practices that promote trust, openness, and collaborative learning, such as implementing agile practices that encourage constructive feedback and open dialogue. In addition, it is essential to train leaders to identify signs of a lack of PS and implement interventions that address interpersonal challenges, such as disagreements or reluctance to admit mistakes. Finally, fostering an organizational culture that values transparency, continuous learning, and mutual support can improve not only team dynamics but also the quality of the software products developed.

\section{Threats to Validity}
\label{sec:threatstovalidity}

The validity of this study is subject to threats that should be considered when interpreting the findings. These threats are discussed as follows.

\textbf{Selection Bias}: Despite following established guidelines for systematic reviews — such as predefined inclusion and exclusion criteria, iterative selection processes, and inter-rater reliability checks using Cohen’s Kappa coefficient — there remains a risk of selection bias. Relevant studies may have been excluded due to variations in search terms, limitations in database indexing, or excluding non-English publications. These factors, inherent in the review process, may have constrained the scope of the literature analyzed.

\textbf{Methodological Limitations}: The methodological framework of this systematic review introduces inherent constraints. Variability in search strategies, inclusion criteria, and data extraction processes can impact the comprehensiveness and reliability of the results. Furthermore, the reliance on quantitative approaches, particularly the widespread use of questionnaires based on Edmondson’s framework, limited the investigation of more context-specific and qualitative aspects of PS. The absence of longitudinal studies further restricted insights into the temporal dynamics of PS and its interactions with different stages of software projects and organizational environments.

\textbf{External validity}: The findings of this review may have limited generalizability across different industries, organizational structures, cultural contexts, and technological settings. The scarcity of studies focusing specifically on PS in software development further constrains the ability to extrapolate results to broader team and organizational contexts.

\textbf{Reliability and Comparability}: The heterogeneity of methodologies across the reviewed studies introduces challenges related to the reliability and comparability of findings. Differences in study designs, theoretical frameworks, and data collection approaches may affect the consistency of synthesized results. While efforts were made to categorize findings systematically, discrepancies in how PS was measured and operationalized in different studies may influence interpretations.

To enhance reliability, we employed structured data extraction procedures and ensured transparency in reporting. However, future research would benefit from more standardized measurement instruments and methodological triangulation to strengthen the robustness of PS assessments in software development.

\textbf{Synthesis-level biases}:
We acknowledge the potential for synthesis-level biases, such as confirmation bias and interpretation bias, during the process of integrating findings across studies. To mitigate these risks, we adopted several strategies. First, we systematically applied a transparent and predefined coding scheme during the thematic analysis, which helped reduce the influence of individual expectations or preferences. Second, multiple researchers were involved in the coding and synthesis process, allowing for cross-validation and critical discussion of interpretations. Discrepancies were resolved through dialogue and consensus, promoting a more balanced and objective integration of findings. Third, we used the exact variable names and labels as defined by the original study authors, ensuring that interpretations remained faithful to the authors’ conceptualizations and avoiding reclassification that could introduce bias. These measures aimed to enhance the credibility and trustworthiness of the synthesis while acknowledging the interpretive nature of the data integration.

Despite its limitations, the systematic approach employed in this study enhances the transparency and replicability of its findings. Future research on PS should seek to address the gaps identified in this study through the adoption of mixed-methods approaches, the execution of longitudinal studies, and the exploration of PS across a more diverse range of SE contexts.

\section{Conclusion}
\label{sec:conclusion}

This paper presents a systematic review of the literature on PS in software development. It includes 28 studies, and their findings were synthesized to answer the defined research questions. These questions explored the dimensions of PS research, including the number of studies per year and country, the PS concepts and instruments used, and antecedents and consequences of PS in software workspaces.

PS emerges as a fundamental factor in the success of software development teams, influencing team dynamics, socio-technical decision-making, and overall software quality. By enabling open discussions about failures, fostering innovation, and supporting continuous learning, PS enhances both team effectiveness and technical excellence. Its benefits, however, require intentional effort, including leadership behaviors that build trust, interventions that raise awareness, and the cultivation of interpersonal trust and learning cultures.

Despite significant progress in understanding PS in software development, key challenges remain. Our review identified a lack of longitudinal studies and a limited focus on specific factors that affect PS. Furthermore, the practical application and evaluation of findings within the industry remain underexplored, restricting the application of academic findings into actionable practices.

This review provides an understanding of PS in software development, offering insights that should guide researchers and practitioners in fostering safer and more collaborative work environments. The findings underscore the importance of PS in enhancing team performance and improving software quality.

Future research can address the gaps identified in this review, including developing intervention strategies tailored to SE teams, identifying contextual elements that affect PS, longitudinal studies to understand the dynamics of PS over time, and investigations into the impact of cultural and organizational diversity on PS. Furthermore, exploring the practical application of PS concepts in industrial settings can provide valuable insights for practitioners. As the field matures and more quantitative studies accumulate, future work should also consider conducting meta-analyses to synthesize effect sizes and assess the overall strength of associations involving psychological safety in software engineering contexts. Lastly, there is a need for a comparative evaluation of PS interventions and their effectiveness across diverse team structures and organizational contexts.

\section*{Acknowledgments}

This study was financed in part by the Coordenação de Aperfeiçoamento de Pessoal de Nível Superior Brasil (CAPES) - Finance Code 001, and by the Conselho Nacional de Desenvolvimento Científico e Tecnológico (CNPq).

\section*{CRediT authorship contribution statement}
\textbf{Beatriz Santana}: Writing – original draft, Visualization, Software, Methodology, Formal analysis, Data curation, Conceptualization. \textbf{Lidivânio Monte}: Writing – review \& editing, Data curation, Formal analysis. \textbf{Bianca Santana de Araújo} : Data curation, Formal analysis. \textbf{Glauco Carneiro}: Writing – review \& editing, Validation. \textbf{Savio Freire}: Writing – review \& editing, Formal analysis, Validation, Supervision. \textbf{José Amancio Macedo Santos}: Writing – review \& editing, Validation, Supervision. \textbf{Manoel Mendonça}: Writing – review \& editing, Validation, Supervision, Conceptualization. 

\section*{Declaration of competing interest}

The authors declare that they have no known financial interests or personal relationships that could have influenced the work presented in this paper. During the preparation of this work, the authors used ChatGPT to assist with writing refinement and language adaptation. After using this tool, they carefully reviewed and edited the content as necessary and took full responsibility for the final publication.

\section*{Supplementary material}
The complete data set is available at https://zenodo.org/records/15436727
\bibliographystyle{elsarticle-harv}

\input{main.bbl}
\pagebreak
\appendix
\input{8.Appendix}
\pagebreak
\input{9.Appendix2}

\end{document}

%% file: 8.Appendix.tex
\section{STUDIES INCLUDED IN THE REVIEW}
\label{appendix:list}
\begin{itemize}
    \item[P1.] Lenberg, P., Feldt, R., 2018. Psychological safety and norm clarity in software engineering teams, in: Proceedings of the 11th international workshop on cooperative and human aspects of software engineering, pp. 79–86.
    \item[P2.] Tkalich, A., Smite, D., Andersen, N.H., Moe, N.B., 2022. What happens to psychological safety when going remote? IEEE Software 41, 113–122.
    \item[P3.] Safdar, U., Badir, Y.F., Afsar, B., 2017. Who can i ask? how psychological safety affects knowledge sourcing among new product development team members. The Journal of High Technology Management Research 28, 79–92.
    \item[P4.] Alami, A., Krancher, O., 2022. How scrum adds value to achieving software quality? Empirical Software Engineering 27, 165.
    \item[P5.] Agren, P., Knoph, E., Berntsson Svensson, R., 2022. Agile software development one year into the covid-19 pandemic. Empirical Software Engineering 27, 121.
    \item[P6.] Khanna, D., Wang, X., 2022. Are your online agile retrospectives psychologically safe? the usage of online tools, in: International Conference on Agile Software Development, Springer. pp. 35–51.
    \item[P7.] Christensen, M.A., Tell, P., 2022. Building a toolbox for working with psychological safety in agile software teams, in: International Conference on Agile Software Development, Springer. pp. 82–98.
    \item[P8.] Gustavsson, T., 2018. Impacts on team performance in large-scale agile software development, in: 2018 Joint of the 17th Business Informatics Research, Workshops and Doctoral Consortium, BIR-WS 2018, 24 September 2018 through 26 September 2018, CEUR-WS. pp. 421–431.
    \item[P9.] Hood, A.C., Bachrach, D.G., Zivnuska, S., Bendoly, E., 2016. Mediating effects of psychological safety in the relationship between team affectivity and transactive memory systems. Journal of Organizational Behavior 37, 416–435.
    \item[P10.] Hennel, P., Rosenkranz, C., 2021. Investigating the “socio” in socio-technical development: The case for psychological safety in agile information systems development. Project Management Journal 52, 11–30.
    \item[P11.] Gustavsson, T., 2022. Team performance in large-scale agile software development, in: Advances in Information Systems Development: Crossing Boundaries Between Development and Operations in Information Systems. Springer, pp. 237–254.
    \item[P12.] Zhang, J., Akhtar, M.N., Zhang, Y., Rofcanin, Y., 2019. High-commitment work systems and employee voice: A multilevel and serial mediation approach inside the black box. Employee Relations: The International Journal 41, 811–827.
    \item[P13.] Alami, A., Zahedi, M., Krancher, O., 2024. The role of psychological safety in promoting software quality in agile teams. Empirical Software Engineering 29, 119.
    \item[P14.] Trinkenreich, B., Gerosa, M.A., Steinmacher, I., 2024. Unraveling the drivers of sense of belonging in software delivery teams: Insights from a large-scale survey, in: Proceedings of the IEEE/ACM 46th International Conference on Software Engineering, pp. 1–12. 
    \item[P15.] Verwijs, C., Russo, D., 2023. The double-edged sword of diversity: How diversity, conflict, and psychological safety impact software teams. IEEE Transactions on Software Engineering.
    \item[P16.] Alami, A., Krancher, O., Paasivaara, M., 2022. The journey to technical excellence in agile software development. Information and Software Technology 150, 106959.
    \item[P17.] Alami, A., Zahedi, M., Krancher, O., 2023. Antecedents of psychological safety in agile software development teams. Information and Software Technology 162, 107267.
    \item[P18.] Barros, L., Tam, C., Varajao, J., 2024. Agile software development projects– unveiling the human-related critical success factors. Information and Software Technology 170, 107432.
    \item[P19.] Gaan, N., Sahoo, A., 2023. Adhocracy culture buffers for mindfulness outcome: A cross-level moderated mediation analysis. Telematics and Informatics Reports 11, 100071.
    \item[P20.] Khan, K., e Habiba, U., Aziz, S., Sabeen, Z., Zeeshan, A., Naz, Z., Waseem, M., 2024. Remote work arrangement: a blessing in disguise for socially anxious individuals. Frontiers in Psychology 14, 1152499.
    \item[P21.] Ahmad, I., Gao, Y., Su, F., Khan, M.K., 2023. Linking ethical leadership to followers’ innovative work behavior in pakistan: the vital roles of psychological safety and proactive personality. European Journal of Innovation Management 26, 755–772.
    \item[P22.] Santana, B.S.D., Freire, S., Cruz, L., Monte, L., Mendonca, M., Santos, J.A.M., 2023. Exploring psychological safety in software engineering: Insights from stack exchange, in: Proceedings of the XXXVII Brazilian Symposium on Software Engineering, pp. 503–513.
    \item[P23.] Santana, B., Freire, S., Santos, J.A.M., Mendon¸ca, M., 2024. Psychological safety in the software work environment. IEEE Software 41, 86–94. doi:10.1109/MS.2024.3386532.
    \item[P24.] Faraj, S., Yan, A., 2009. Boundary work in knowledge teams. Journal of applied psychology 94, 604.
    \item[P25.] Kakar, A.K., 2018. How do team cohesion and psychological safety impact knowledge sharing in software development projects? Knowledge and Process Management 25, 258–267.
    \item[P26.] Thorgren, S., Caiman, E., 2019. The role of psychological safety in implementing agile methods across cultures. Research-Technology Management 62, 31–39.
    \item[P27.] Buvik, M.P., Tkalich, A., 2022. Psychological safety in agile software development teams: Work design antecedents and performance consequences. in Proceedings of the 55th Hawaii International Conference on System Sciences.
    \item[P28.] Ahmad, M.O., 2023. Psychological safety, leadership and non-technical debt in large-scale agile software development, in: 2023 18th Conference on Computer Science and Intelligence Systems, IEEE. pp. 327–334.    
\end{itemize}

%% file: 9.Appendix2.tex
\section{CATEGORIES OF FACTORS PER STUDY}

\begin{table}[H]
\centering
\caption{Categories of factors per study}
\label{tab:appendix_Factors_Study}
\resizebox{\columnwidth}{!}{%
\begin{tabular}{|ll|r|l|l|}
\hline
\multicolumn{1}{|l|}{\multirow{18}{*}{\textbf{\begin{tabular}[c]{@{}l@{}}Antecedents\\ of PS\end{tabular}}}} & \textbf{Level} & \multicolumn{1}{l|}{\textbf{ID}} & \textbf{Study} & Factor \\ \cline{2-5} 
\multicolumn{1}{|l|}{} & \multirow{4}{*}{Individual} & 6 & ~\citep{06} & Behaviors \\ \cline{3-5} 
\multicolumn{1}{|l|}{} &  & 22 & ~\citep{22} & \begin{tabular}[c]{@{}l@{}}Indicators of lack of PS\end{tabular} \\ \cline{3-5} 
\multicolumn{1}{|l|}{} &  & 23 & ~\citep{23} & Interpersonal challenges \\ \cline{3-5} 
\multicolumn{1}{|l|}{} &  & 17 & ~\citep{17} & Openness, no blame \\ \cline{2-5} 
\multicolumn{1}{|l|}{} & \multirow{7}{*}{Team} & 2 & ~\citep{02} & Spontaneous interaction \\ \cline{3-5} 
\multicolumn{1}{|l|}{} &  & 5 & ~\citep{05} & Social interactions \\ \cline{3-5} 
\multicolumn{1}{|l|}{} &  & 17 & ~\citep{17} & Collective decision-making \\ \cline{3-5} 
\multicolumn{1}{|l|}{} &  & 9 & ~\citep{09} & \begin{tabular}[c]{@{}l@{}}Transactive memory systems, \\ group affectivity\end{tabular} \\ \cline{3-5} 
\multicolumn{1}{|l|}{} &  & 21 & ~\citep{21} & \begin{tabular}[c]{@{}l@{}} Trusting and respectful \\ interpersonal relationships \end{tabular} \\ \cline{3-5} 
\multicolumn{1}{|l|}{} &  & 10 & ~\citep{10} & \begin{tabular}[c]{@{}l@{}}Social agile  practices\end{tabular} \\ \cline{3-5} 
\multicolumn{1}{|l|}{} &  & 27 & ~\citep{27} & \begin{tabular}[c]{@{}l@{}}Team autonomy,\\  team performance\end{tabular} \\ \cline{2-5} 
\multicolumn{1}{|l|}{} & \multirow{7}{*}{Organizational} & 12 & ~\citep{12} & \begin{tabular}[c]{@{}l@{}}High-commitment \\ work systems\end{tabular} \\ \cline{3-5} 
\multicolumn{1}{|l|}{} &  & 7 & ~\citep{07} & Tools \\ \cline{3-5} 
\multicolumn{1}{|l|}{} &  & 24 & ~\citep{24} & Boundary work \\ \cline{3-5} 
\multicolumn{1}{|l|}{} &  & 26 & ~\citep{26} & Cultural differences \\ \cline{3-5} 
\multicolumn{1}{|l|}{} &  & 17 & ~\citep{17} & Ownership of PS  \\ \cline{3-5} 
\multicolumn{1}{|l|}{} &  & 21 & ~\citep{21} &  \begin{tabular}[c]{@{}l@{}}Leader behavior, \\designing a team for learning \end{tabular}  \\ \hline
\multicolumn{1}{|l|}{\multirow{16}{*}{\textbf{\begin{tabular}[c]{@{}l@{}}Consequences\\  of PS\end{tabular}}}} & \multirow{2}{*}{Individual} & 1 & ~\citep{01} & \begin{tabular}[c]{@{}l@{}}Individual job satisfaction\end{tabular} \\ \cline{3-5} 
\multicolumn{1}{|l|}{} &  & 3 & ~\citep{03} & \begin{tabular}[c]{@{}l@{}}Knowledge sourcing\end{tabular} \\ \cline{2-5} 
\multicolumn{1}{|l|}{} & \multirow{14}{*}{Team} & 1 & ~\citep{01} & Team performance \\ \cline{3-5} 
\multicolumn{1}{|l|}{} &  & 8 & ~\citep{08} & Team performance \\ \cline{3-5} 
\multicolumn{1}{|l|}{} &  & 11 & ~\citep{11} & Team performance \\ \cline{3-5} 
\multicolumn{1}{|l|}{} &  & 14 & ~\citep{14} & \begin{tabular}[c]{@{}l@{}}Belonging to the team\end{tabular} \\ \cline{3-5} 
\multicolumn{1}{|l|}{} &  & 15 & ~\citep{15} & Team effectiveness \\ \cline{3-5} 
\multicolumn{1}{|l|}{} &  & 18 & ~\citep{18} & Team capabilities \\ \cline{3-5} 
\multicolumn{1}{|l|}{} &  & 27 & ~\citep{27} & \begin{tabular}[c]{@{}l@{}}Team reflexivity,\\  team performance\end{tabular} \\ \cline{3-5} 
\multicolumn{1}{|l|}{} &  & 9 & ~\citep{09} & Transactive memory systems \\ \cline{3-5} 
\multicolumn{1}{|l|}{} &  & 25 & ~\citep{25} & Knowledge sharing \\ \cline{3-5} 
\multicolumn{1}{|l|}{} & & 4 & ~\citep{04} & Technical excellence \\ \cline{3-5} 
\multicolumn{1}{|l|}{} &  & 13 & ~\citep{13} & Software quality \\ \cline{3-5} 
\multicolumn{1}{|l|}{} &  & 16 & ~\citep{16} & Software quality \\ \cline{3-5} 
\multicolumn{1}{|l|}{} &  & 18 & ~\citep{18} & Customer involvement \\ \cline{3-5} 
\multicolumn{1}{|l|}{} &  & 10 & ~\citep{10} & Social agile practices \\ \hline
\multirow{5}{*}{\textbf{\begin{tabular}[c]{@{}l@{}}Other\\  findings\end{tabular}}} & \multirow{5}{*}{} & 19 & ~\citep{19} & \begin{tabular}[c]{@{}l@{}}Mindfulness and \\ team performance\end{tabular} \\ \cline{3-5} 
 &  & 20 & ~\citep{20} & \begin{tabular}[c]{@{}l@{}}Social anxiety disorder \\ and psychological distance\end{tabular} \\ \cline{3-5} 
 &  & 28 & ~\citep{28} & \begin{tabular}[c]{@{}l@{}}Ethical leadership and \\ innovative work behavior\end{tabular} \\ \cline{3-5} 
 &  & 1 & ~\citep{01} & Team norms \\ \cline{3-5} 
 &  & 21 & ~\citep{21} & Brain drain \\ \hline
\end{tabular}%
}
\end{table}